\documentclass[prl,twocolumn,showpacs]{revtex4}
\usepackage{graphicx}
\begin{document}

\title{Orbital ordering in frustrated Jahn-Teller systems}
\author{M. V. Mostovoy and D. I. Khomskii}
\affiliation{Materials Science Center, University of
Groningen, Nijenborgh 4, 9747 AG Groningen,
  The Netherlands}

\begin{abstract}

We consider the superexchange in `frustrated' Jahn-Teller
systems, such as the transition metal oxides NaNiO$_2$,
LiNiO$_2$, and ZnMn$_2$O$_4$, in which transition metal
ions with doubly degenerate orbitals form a triangular or
pyrochlore lattice and are connected by the $90^\circ$
metal-oxygen-metal bonds. We show that this interaction is
much different from a more familiar exchange in systems
with the $180^\circ$-bonds, e.g. perovskites. In contrast
to the strong interplay between the orbital and spin
degrees of freedom in perovskites, in the
$90^\circ$-exchange systems spins and orbitals are
decoupled: the spin exchange is much weaker than the
orbital one and it is ferromagnetic for all orbital
states. Due to frustration, the mean-field orbital ground
state is strongly degenerate. Quantum orbital fluctuations
select particular ferro-orbital states, such as the one
observed in NaNiO$_2$. We also discuss why LiNiO$_2$ may
still behave as an orbital liquid.

\end{abstract}

\date{\today}


\maketitle

There is a large class of compounds containing transition
metal (TM) ions with orbital degeneracy - the so-called
Jahn-Teller (JT) systems. Orbital degrees of freedom and,
in particular, their ordering or the cooperative
Jahn-Teller effect, give rise to a very rich physics
\cite{KugelKhomskii,TokuraNagaosa}. The interplay between
orbitals and spins often leads to rather peculiar magnetic
structures, turning e.g. cubic perovskites like KCuF$_3$
into quasi-one-dimensional spin-chain materials
\cite{KugelKhomskii}. It was recently suggested that in
some JT systems  orbitals remain disordered at all
temperatures, forming an orbital liquid state
\cite{Ishihara,FeinerOlesZaanen,Keimer,KhaliullinMaekawa}.
The suppression of an orbital ordering was discussed for
materials with a simple cubic structure, e.g. the
perovskite LaTiO$_3$, and was attributed to a strong
anisotropy of orbital interactions and a strong coupling
between orbital and spin excitations. One would expect to
find an even stronger tendency to form an orbital liquid
in JT compounds with more complicated crystal structures,
especially, in `geometrically frustrated' systems
\cite{Ramirez}, e.g. the spinel ZnMn$_2$O$_4$, in which
Mn$^{3+}$ ions with the configuration ($t^3_{2g}e^1_g$)
form a pyrochlore lattice of corner sharing tetrahedra, or
the layered materials LiNiO$_2$ and NaNiO$_2$ with a
triangular lattice of low-spin  Ni$^{3+}$ ion in the
($t^6_{2g}e^1_g$) configuration. Orbital ordering in such
systems is an open issue, interesting from both
theoretical and practical points of view, as some of these
compounds (LiNiO$_2$, NaNiO$_2$) are now studied as
promising materials for rechargeable batteries, and their
performance may be significantly affected by the JT
effect.

An important difference between the TM oxides with
`frustrated lattices' and perovskites is the angle between
the oxygen-metal bonds connecting two neighboring TM ions.
While in perovskites this angle is close to $180^\circ$,
in the geometrically frustrated JT systems it is typically
$90^\circ$. Though usually ignored, this difference has
important consequences for orbital and magnetic orderings.
In this Letter we derive the Hamiltonian of the
$90^\circ$-exchange and show that, unlike in perovskites,
in frustrated JT systems orbitals and spins are
essentially decoupled and that orbital and magnetic
orderings can be considered separately from each other.
Considering in particular layered materials with a
triangular lattice, we show that the orbital interactions
are strongly frustrated, which results in a large number
of disordered mean-field ground states. However, we argue
that this frustration is lifted by quantum orbital
fluctuations that stabilize a ferro-orbital state and
induce a gap in the spectrum of orbital excitations. We
also show that the ordering of spins in the layers is
ferromagnetic. These results agree with experimental
properties of NaNiO$_2$ \cite{Chappel} and lead us to a
conclusion that the puzzling absence of both spin and
orbital ordering in LiNiO$_2$ \cite{Kitaoka,Reynaud}
results from disorder rather than from the superexchange.

We first obtain the exchange Hamiltonian describing
interactions between TM ions with one electron or hole on
doubly degenerate $e_g$-levels. In this case orbital
states are conveniently described by introducing isospins
(or pseudospins)  ${\bf T}_j$ on each TM site $j$, which
act on the up and down states, $\left|\right.T^z = \pm
\frac{1}{2}\rangle$ identified with, respectively,
$d_{3z^2 - r^2}$ and $d_{x^2 - y^2}$ orbitals. Quite
generally, an effective interaction between the orbitals
and spins on pairs of neighboring TM sites $i$ and $j$ has
the form \cite{KugelKhomskii}
\begin{equation}
H_{ST} = \sum_{ij} \left[J_{S} ({\bf S}_i {\bf S}_j) + J_T
T_i T_j + J_{ST} ({\bf S}_i {\bf S}_j) T_i T_j\right]
\label{HKK}
\end{equation}
(for simplicity, we do not show here the full structure of
the orbital interactions, which are strongly anisotropic
both in real and isospin space).  In perovskites with the
$180^\circ$-bonds the spin- and orbital-exchange constants
$J_S$ and $J_T$, as well as the constant $J_{ST}$,
describing the coupling between orbitals and spins, are
all of the same order of magnitude. This makes the spin
and orbital orderings dependent on each other
\cite{KugelKhomskii}. The strong interplay between
orbitals and spins is also crucial for the mechanisms of
suppression of these orderings used in
Refs.~\cite{FeinerOlesZaanen,Li,Mila} to explain an
apparent absence of both orbital and magnetic ordering in
LiNiO$_2$ \cite{Kitaoka,Reynaud}. In particular, in the
$SU(4)$ version of the model Eq.(\ref{HKK}) \cite{Li,Mila}
both the spin and orbital interactions are isotropic and
$J_S = J_T = \frac{1}{4} J_{ST}$. We first show that the
actual situation in the $90^\circ$-systems is quite
different: the orbital exchange is strongly anisotropic
and $J_S \ll J_{ST} \ll J_T$.

\begin{figure}
\centering
\includegraphics[width=7cm]{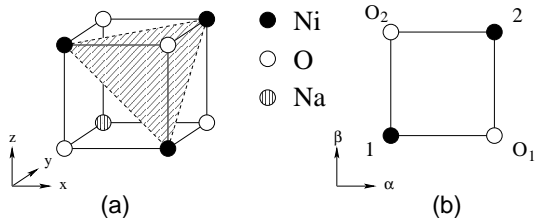}
\caption{\label{nio} The crystal structure of NaNiO$_2$
(a) and a plaquet\-te in the $\alpha\beta$-plane
($\alpha,\beta = x,y,z$) formed by two nearest-neighbor Ni
ions, $1$ and $2$, and two oxygens, O$_1$ and O$_2$ (b).}
\end{figure}

We derive the exchange Hamiltonian using as an example the
layered material NaNiO$_2$, in which Ni and Na ions occupy
consecutive $[111]$-planes of the NaCl structure (see
Fig.~\ref{nio}(a)). In each plane the low-spin Ni$^{3+}$
ions with one electron on the doubly degenerate $e_g$
level form a triangular lattice. The exchange between two
nearest-neighbor Ni ions in the planes occurs via common
oxygens and the angle between the two oxygen $p$-orbitals
involved in this exchange is $90^\circ$. The exchange
between Ni ions from two neighboring planes (separated by
a plane of Na ions) is weak and will be neglected here.

The electron hopping in the $z$-direction can only occur
between the $p_z$ oxygen and the $d_{3z^2 - r^2}$ nickel
orbitals. The corresponding hopping amplitude is denoted
by $t$. All other processes are forbidden by symmetries of
the oxygen $p$ and nickel $e_g$ orbitals. Thus, the
hopping amplitude in the $z$-direction is $t$ for the
$|T^z = +\frac{1}{2}\rangle$ nickel state, and $0$ for the
$|T^z = -\frac{1}{2}\rangle$ state. More generally, the
hopping in the $\alpha$-direction ($\alpha = x,y,z$) is
only nonzero for the oxygen $p_\alpha$ and the nickel
$d_{3\alpha^2 - r^2}$ orbitals. It is convenient to
introduce the isospin operators $I^{\alpha} = {\bf T}
\cdot {\bf n^{\alpha}}$, where ${\bf n}^\alpha$ are three
unit vectors in the $(T^x,T^z)$ plane
\begin{equation}\label{I}
{\bf n}^x  =  - \frac{1}{2} {\hat {\bf z}} -
\frac{\sqrt{3}}{2} {\hat {\bf x}}, \,\,\, {\bf n}^y  =
-\frac{1}{2} {\hat {\bf z}} + \frac{\sqrt{3}}{2} {\hat
{\bf x}}, \,\,\, {\bf n}^z  =  {\hat {\bf z}}.
\end{equation}
The operators $I^x$ and $I^y$ play the role of $T^z$ for
the $x$ and $y$-directions, i.e. they describe the
occupation of, respectively, the $d_{3x^2 - r^2}$ and
$d_{3y^2 - r^2}$ orbital. The hopping amplitude in the
$\alpha$-direction is then $t$ for the $|I^{\alpha} = +
\frac12\rangle$ state and $0$ for the $|I^\alpha = -
\frac12\rangle$ state.

Figure~\ref{nio}(b) shows a plaquette in the
$\alpha\beta$-plane ($\alpha,\beta = x,y,z$) formed by two
neighboring Ni sites $1$ and $2$ and the two oxygens
shared by the oxygen octahedra surrounding the TM ions. An
effective exchange Hamiltonian for two $e_g$ nickel
electrons is obtained by expanding the plaquette energy in
powers of the hopping amplitude $t$, which is assumed to
be much smaller than the energy $\Delta$ of the electron
transfer from O to Ni and the Hubbard repulsion on oxygen
$U_p$. The orbital and spin exchange appears in the fourth
order of the perturbative expansion and the higher-order
terms are neglected here.

We first neglect also the Hund's rule coupling between the
spins of two holes on oxygens. In that case the
$90^\circ$-exchange is independent of a spin configuration
and only involves orbitals. The orbital exchange
Hamiltonian of the plaquette in the $\alpha\beta$-plane
has the form
\begin{equation}
H_{T} = J_T \,{\cal I}^{\,\alpha\beta} + const,\label{HT}
\end{equation}
where
\begin{equation}
{\cal I}^{\,\alpha\beta} = \left(\frac{3}{2} -
I_1^\alpha\right)\! \left(\frac{3}{2} - I_2^\beta\right)\!
+\! \left(\frac{3}{2} - I_1^\beta\right)\!
\left(\frac{3}{2} - I_2^\alpha\right)\label{Iab}
\end{equation}
and $J_T$ is the coupling constant specified below. The
factor $\frac{3}{2} - I_1^\alpha$ counts the number of the
oxygen electrons that can hop to the Ni site $1$ along the
bond in the $\alpha$-direction. This number equals $1$, if
the $e_g$-electron is in the state $|I_1^\alpha =
+\frac{1}{2}\rangle$, and $2$, if the electron occupies
the state $|I_1^\alpha = -\frac{1}{2}\rangle$. (For one
hole on $e_g$-level, as in Cu$^{2+}$, the exchange
interaction corresponding to Eq.(\ref{Iab}) involves
combinations $\left(\frac{1}{2} + I^\alpha\right)$ instead
of $\left(\frac{3}{2} - I^\alpha\right)$.) The two terms
in Eq.(\ref{Iab}) correspond to the two exchange paths on
the plaquette through the oxygens O$_1$ and O$_2$.

The orbital coupling
\begin{equation}
J_T =-\frac{4t^2}{\Delta^2(2\Delta+U_p)} + \frac{2
t^4}{\Delta^3} = \frac{2 t^4 U_p}{\Delta^3 (2\Delta +
U_p)} \label{eq:JT}
\end{equation}
is the sum of the contributions of two different
mechanisms of the orbital exchange. The first term in
Eq.(\ref{eq:JT}) is due to the exchange via the
intermediate states with two holes on one of the oxygens
from the plaquette. The mechanism resulting in the second
term in Eq.(\ref{eq:JT})  works even for infinite $U_p$,
when the two-hole oxygen states are forbidden. To
understand its origin consider first a single Ni site
inside the oxygen octahedron. The hopping of electrons
from the $6$ oxygens to nickel results in the energy
decrease $\delta E$. For two neighboring Ni sites this
decrease is smaller than $2\,\delta E$, as the surrounding
octahedra share two oxygens and the two-hole oxygen states
are forbidden for infinite $U_p$. The effect of the
blocking of the two-hole states on common oxygens depends
on orbital states of the $e_g$-electrons, which gives rise
to the orbital exchange with the coupling constant
$\frac{2 t^4}{\Delta^3}$.

The intermediate states with two holes on one oxygen also
result in a spin exchange. As in such states the holes
occupy two mutually orthogonal orbitals $p_\alpha$ and
$p_\beta$, the spin exchange is only possible due to the
Hund's rule coupling $J_H$ between the spins of the holes
(in systems with the $180^\circ$-bonds the spin exchange
occurs even for $J_H = 0$). The corresponding exchange
Hamiltonian involves both spin and isospin operators and
to the lowest-order in powers of $J_H$ it has the form
\begin{equation}
H_{TS} = - J_{TS} {\cal I}^{\,\alpha\beta}
\left[\frac{3}{4} + ({\bf S}_1 {\bf S}_2)\right],
\label{eq:HTS}
\end{equation}
where $J_{TS} = \frac{4t^4 J_H}{\Delta^2 (2\Delta +
U_p)^2}$ and ${\cal I^{\,\alpha\beta}}$ is given by
Eq.(\ref{Iab}). The interaction described by
Eq.(\ref{eq:HTS}) is different the $180^\circ$-exchange in
two important respects. First, since for any orbital state
the expectation value $\langle {\cal
I}^{\,\alpha\beta}\rangle>0$, the spin exchange is
effectively ferromagnetic independent of an orbital state
(the average spin-exchange coupling $- J_{TS} \langle
{\cal I}^{\,\alpha\beta}\rangle<0$ is negative). Second,
since $J_H \ll U_p$, the spin-isospin coupling in the
$90^\circ$-systems is weaker than the pure orbital
exchange, described by Eq.(\ref{HT}):
\[
\frac{J_{TS}}{J_T} = \frac{J_H}{U_p}\,
\frac{2\Delta}{(2\Delta + U_p)}.
\]
Thus, in $90^\circ$-exchange systems spins and orbitals
cannot strongly influence each other, i.e. they are
essentially decoupled. (A similar conclusion was reached
by Reitsma and Feiner (unpublished). We are grateful to L.
F. Feiner for informing us about their results.)

\begin{figure}
\includegraphics[width=4cm]{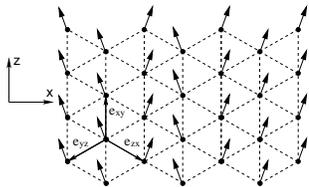}
\caption{\label{triang} The triangular lattice formed by
Ni ions in the $[111]$-plane. Shown also is a disordered
mean-field ground state, in which the isospins form lines
parallel to the unit vector ${\bf e}_{xy}$, such that
$\langle T^z_j \rangle$ is the same on all lattice sites,
while the sign of $\langle T^x_j \rangle$ varies arbitrary
from line to line.}
\end{figure}

We can now obtain the Hamiltonian describing the orbital
interactions in the Ni layers of NaNiO$_2$ (see
Fig.~\ref{nio}a).  In each layer Ni ions form a triangular
lattice, the sites of which lie on intersections of the
three sets of lines parallel to the unit vectors ${\bf
e}_{xy}$, ${\bf e}_{yz}$, and ${\bf e}_{zx}$, as shown in
Fig.~\ref{triang}. It will be convenient to identify ${\bf
e}_{\alpha \beta}$ with the three unit vectors in the
isospace (see  Eq.(\ref{I})) by ${\bf n}^x = {\bf
e}_{yz}$, ${\bf n}^y = {\bf e}_{zx}$, and ${\bf n}^z =
{\bf e}_{xy}$. The bonds between the nearest-neighbor
sites $j$ and $j + e_{\alpha\beta}$ of the triangular
lattice (the lattice constant is put to $1$) are diagonals
of the Ni-O plaquettes lying in the $\alpha\beta$ plane
(cf. Fig.~\ref{nio}b). Hence, the Hamiltonian, describing
orbital interactions on the triangular lattice has the
form:
\begin{equation}\label{Htau}
H_{T} = J_T \sum_j\sum_{\stackrel{\scriptstyle
\alpha\beta} {\scriptstyle \alpha\neq\beta}}I_j^\alpha
I^\beta_{j+ e_{\alpha\beta}}.
\end{equation}
The terms linear in the operators $I^\alpha_j$ (see
Eq.(\ref{HT})) are canceled in (\ref{Htau}), since
$\sum_\alpha I^\alpha_j = 0$.

The Hamiltonian Eq.(\ref{Htau}) is invariant under the
global rotation of the isospin operators $T^\alpha$ over
the angle $\frac{2\pi}{3}$ around the $y$-axis combined
with the rotation of the triangular lattice over the same
angle. This transformation is equivalent to a cyclic
permutation of the indices $x$, $y$, and $z$ of the
vectors ${\bf e}_{\alpha\beta}$ and the isospin operators
$I^\alpha$. Thus the Hamiltonian Eq.(\ref{Htau}) has the
same symmetry as the so-called `compass' model, describing
exchange interactions between TM ions on a cubic lattice
\cite{KugelKhomskii,GKhaliullin} (note, however, that in
our case the three operators $I^x$, $I^y$, and $I^z$ are
not independent).

The orbital exchange on a triangular lattice  turns out to
be strongly frustrated and the mean-field ground state is
not unique. The simplest states with the minimal energy
$E_0 = -\frac{3}{8}J$ per site are ferro-orbital states,
in which $\langle {\bf T}_j \rangle = T {\bf m}$ on all
lattice sites, where ${\bf m}$ is an arbitrary unit vector
in the $(T^x,T^z)$-plane and $T = \frac{1}{2}$. The energy
of these states is independent of orientation of ${\bf
m}$, even though the Hamiltonian (\ref{Htau}) is not
invariant under arbitrary rotations in the
$(T^x,T^z)$-plane  (this continuous ground-state
degeneracy is similar to the one of the `compass' model
\cite{GKhaliullin}). Furthermore, there exist also
disordered mean-field ground states, which can be obtained
from the ferro-orbital states by inverting the sign of
$\langle T^x_j \rangle$ on an arbitrarily selected set of
lines parallel to ${\bf e}_{xy}$ (see Fig.~\ref{triang}).
Such states are ordered along the $xy$-lines, but there
are no long-range correlations between the $x$-projections
of isospins in the transverse direction. By circular
permutations of the $x$, $y$, and $z$ indices one can
obtain similar states, which are only ordered along the
lines parallel to the ${\bf e}_{yz}$ and ${\bf e}_{zx}$
vectors.

\begin{figure}
\includegraphics[width=4cm]{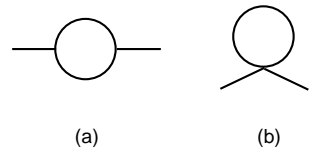}
\caption{\label{diagrams} The self-energy diagrams due to
the qubic (a) and the quartic (b) interactions that result
in a gap opening and a two-dimensional dispersion.}
\end{figure}

This large ground-state degeneracy is lifted by quantum
orbital fluctuations (the so-called `order-from-disorder'
mechanism \cite{Villain,Shender}). First, we have checked
numerically that disordered states (see Fig.~\ref{triang})
have a higher energy of the zero-point fluctuations than
the corresponding uniform states. Second, the zero-point
energy of uniform states has $6$  minima at the isospin
orientations ${\bf m} = \pm {\bf n}^x, \pm {\bf n}^y, \pm
{\bf n}^z$. In the original notation these are the
ferro-orbital states with one of the orbitals $d_{3z^2 -
r^2}$, $d_{3x^2 - r^2}$, $d_{3y^2 - r^2}$, $d_{x^2 -
y^2}$, $d_{y^2 - z^2}$, and $d_{z^2 - x^2}$ occupied at
each site. Furthermore, the anharmonicity effects
\cite{KhomskiivdBrink} usually stabilize elongated
octahedra, i.e. the orbitals $d_{3z^2 - r^2}$, $d_{3x^2 -
r^2}$, or $d_{3y^2 - r^2}$. As was mentioned above, spins
in layers order ferromagnetically.

We would like to point out that the standard description
of orbital excitations with non-interacting bosons
(`isospin-wave approximation') is insufficient for the
calculation of quantum corrections to the ground-state
energy of the frustrated Hamiltonian Eq.(\ref{Htau}). In
this approximation the orbital excitation spectrum for the
six ground states is one-dimensional
\begin{equation}
\omega_{\bf q} = 3 \sqrt{2} T J_T  \left| \sin\frac{({\bf
q}\cdot {\bf m})}{2} \right|, \label{eq:1d}
\end{equation}
which is a direct consequence of the absence of long-range
correlations between chains in disordered mean-field
ground states. The gapless one-dimensional spectrum leads
to infrared-divergent fluctuations. To get rid of them,
one has to take into account interactions between the
bosons, which can be consistently done in the large
isospin limit, $T \gg 1$. The interactions suppress
quantum fluctuations by opening a gap and inducing a
dispersion in the direction transverse to ${\bf m}$. This
can be understood from the fact that the self-energy
diagrams, shown in Fig.~\ref{diagrams} (a) and (b),
diverge unless a gap is introduced (in field theory a
similar mechanism is known as a `dynamical mass
generation' \cite{Coleman}). The physical origin of the
gap is the breaking of the continuous ground-state
degeneracy by quantum fluctuations \cite{GKhaliullin}. In
the large-$T$ limit the quantum fluctuations are
relatively small and the gap $\Delta \propto J_T \sqrt{T}$
is much smaller than the band width $W \sim 3 \sqrt{2} J_T
T$ of the orbital excitations. The dispersion in the
transverse direction is of the order of $\Delta$
\cite{details}.

The main conclusions of our theory of the exchange in
frustrated systems (spins are coupled much weaker than
orbitals, the ground state is ordered both ferro-orbitally
and ferromagnetically) are in agreement with the orbital
and magnetic structure of the layered compound NaNiO$_2$,
which undergoes two transitions: At $T_o = 480$K the
oxygen octahedra become elongated, which corresponds to
the ferro-orbital ordering of the $d_{3z^2 - r^2}$-type,
and at the much lower temperature, $T_s = 20$K, the Ni
spins in the $[111]$ layers order ferromagnetically
\cite{Chappel}.

It is, therefore, very puzzling that the structurally
identical material LiNiO$_2$ shows neither orbital nor
spin ordering. In principle, we cannot exclude that the
quantum orbital fluctuations, which for $T = \frac{1}{2}$
are relatively large, destroy the long-range ferro-orbital
ordering and stabilize an orbital liquid (such quantum
melting of the `order-from-disorder' was discussed in the
context of another frustrated system in Ref.
\cite{Lecheminant}). Since for $T = \frac{1}{2}$ there is
no small parameter in the problem, the question whether
the ground state of the orbital Hamiltonian
Eq.(\ref{Htau}) is ordered or disordered, can only be
resolved by numerical calculations that are beyond the
scope of this Letter. In any case, this possible
explanation of the absence of orbital ordering in
LiNiO$_2$ is fundamentally different from those based on a
strong interplay between orbitals and spins
\cite{FeinerOlesZaanen,Li,Mila}. Actually, due to the
decoupling of spins from orbitals in frustrated systems,
an orbital liquid should still have a ferromagnetic order.

We think that the puzzles surrounding LiNiO$_2$ indicate
an importance of electron-lattice interactions and
disorder, not included in the present theory. As was
argued in Refs.~\cite{Chappel,Mertz}, the `intrinsic'
disorder caused by the presence of magnetic Ni ions in the
Li $[111]$ planes, results in a strong interlayer coupling
that frustrates the ferromagnetic spin ordering in Ni
layers. In NaNiO$_2$ such disorder should be weaker: as
the size difference of Na and Ni ions is larger than that
of Li and Ni ions, the alternating stacking of the Na and
Ni $[111]$ layers is more perfect than that of Li and Ni
in LiNiO$_2$. The larger charge radius of the Na ion also
implies stronger electron-lattice interactions in
NaNiO$_2$. In general, such interactions suppress quantum
orbital fluctuations and stabilize a ferro-orbital
ordering \cite{Kanamori,EnglmanHalperin}. This may explain
why NaNiO$_2$ is orbitally ordered, while LiNiO$_2$ is
not.

In conclusion, we developed a theory of the orbital and
spin exchange in JT systems with doubly degenerate
electronic orbitals and $90^\circ$ metal-oxygen-metal
bonds. We showed that in these systems spins and orbitals
are essentially decoupled:  spins interact much weaker
than orbitals and the spin exchange is ferromagnetic for
any orbital state. The orbital exchange between TM ions on
a triangular lattice is strongly frustrated, resulting in
infinite number of disordered classical ground states. An
even stronger frustration can be shown to occur in
spinels, in which TM ions form a pyrochlore lattice (to be
published elsewhere). We showed, however, that small
quantum orbital fluctuations remove the frustration and
induce a ferro-orbital ordering. Our theory is an
agreement with the orbital and magnetic structure observed
in NaNiO$_2$. We also argued that large quantum
fluctuations and `intrinsic' disorder may, in principle,
turn a frustrated system into an orbital liquid, which may
explain the absence of orbital ordering in LiNiO$_2$.

We are grateful to A. Stepanov and to G. Chouteau and
members of his group for extensive discussions of the
experimental situation in LiNiO$_2$ and NaNiO$_2$, and to
G. Khaliullin and L.-F Feiner for the useful discussions.


\begin{thebibliography}{99}
\bibitem{details} Details of the calculation of the
dispersion of orbital excitations, which also requires a
vertex renormalization, will be published elsewhere.
\end{thebibliography}
\end{document}